# Anisotropic penetration depth and optical sum rule violation in $La_{2-x}Sr_xCuO_4$

F. Marsiglio[a] and J.E. Hirsch[b]

[a]Department of Physics, University of Alberta, Edmonton, Alberta, Canada T6G2J1
[b]Department of Physics, University of California, San Diego, La Jolla, CA 92093-0319, USA

We calculate the doping dependence of the penetration depths and optical sum rule violation in $La_{2-x}Sr_xCuO_4$ within the model of hole superconductivity. In the clean limit the predicted in-plane sum rule violation is larger than the c-axis one, however in the presence of disorder the latter one becomes very substantially enhanced.

The optical sum rule violation recently observed by Basov[1] for interplane transport is predicted to occur also for in-plane transport within the model of hole superconductivity[2,3]. Here we calculate the expected magnitude of the effects in both directions as function of doping for $La_{2-x}Sr_xCuO_4$. Recently we discussed this question for $Tl_2Ba_2CuO_y$ [4].

With a reasonable on-site repulsion U=5eV, other parameters in the model are chosen to match certain experimental observations. The maximum $T_c$=37.5K and the reported[5] condensation energy $\epsilon_c$=13μeV per $CuO_2$ unit for a slightly underdoped sample ($T_c$=32K) determine the bandwidth as $D_h$~1.2eV. For two representative cases of nearest neighbor repulsion V=0 and 0.5eV the model requires a correlated hopping term $\Delta t$=0.27eV and $\Delta t$=0.46eV respectively[2]. We choose the anisotropy in hopping amplitudes from band structure estimates, $t_a/t_c$=25.

The calculated monotonic decrease of penetration depths with doping shown in Fig. 1 qualitatively matches observations[6], but the observed c-axis penetration depth is substantially larger than the calculated one. This points to the important role of disorder in c-axis transport[7]. Within BCS theory the penetration depth in the dirty limit is increased to

$$\frac{1}{\lambda_c^2} = \frac{1}{(\lambda_c^{clean})^2} \frac{\pi\Delta}{\hbar/\tau_c} \equiv \frac{1}{(\lambda_c^{clean})^2} \times p \qquad (1)$$

with $\Delta$ the energy gap and $1/\tau_c$ the scattering rate in the c direction. Comparison of calculated and measured anisotropies in penetration depths, shown in Fig. 2, indicates that the effect of disorder in c-axis transport is becoming smaller with doping. Indeed, Uchida et al[6] conclude from analysis of optical spectra that the c-axis scattering rate decreases substantially with doping. For in-plane transport we assume the clean limit for all doping.

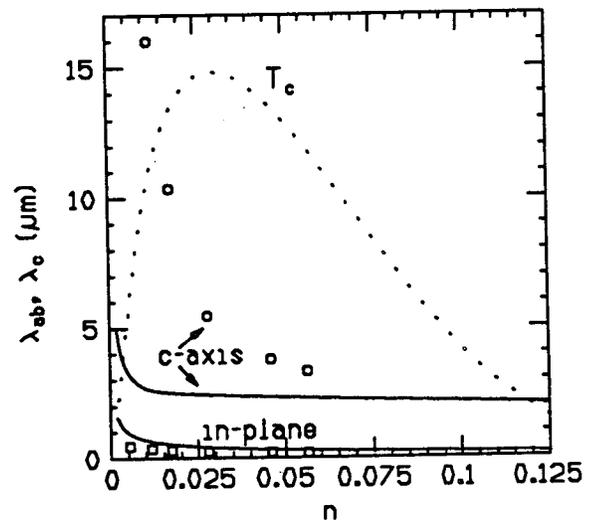

Fig. 1: Penetration depths in a and c directions in the clean limit and experimental values[6] (squares and circles), and calculated $T_c$ versus doping (arb. units). V=0.5eV.

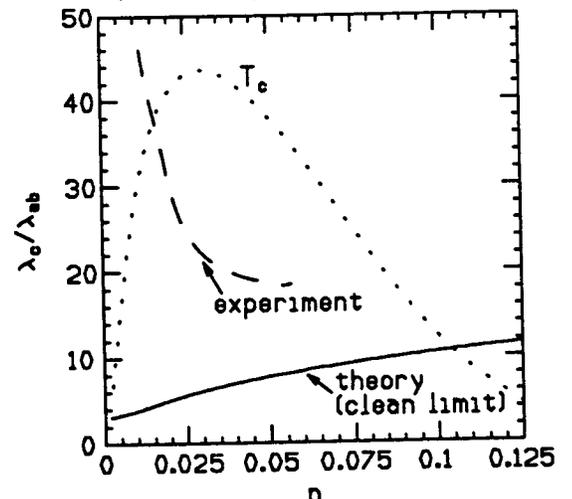

Fig. 2: Anisotropy in penetration depth, calculated in clean limit (solid line) and experimental (dashed line).

The experimental $\lambda_{ab}$ is somewhat smaller than the calculated one, possibly due to contribution of carriers from another band to the superfluid density.

The frequency-dependent conductivity in the superconducting state is given by $\sigma_{1s}(\omega)=D\delta(\omega)+\sigma_{1s}^{reg}(\omega)$. The superfluid weight $D=c^2/8\lambda^2$ in the $\mu$ direction (a or c) has contributions from low and high frequency 'missing areas' in the optical conductivity[3],

$$D_\mu = \delta A_l^\mu + \delta A_h^\mu \qquad (2)$$

with the high frequency missing area $\delta A_h^\mu$ proportional to the lowering of kinetic energy that drives superconductivity[2]. The degree of sum rule violation

$$V_\mu = \frac{\delta A_h^\mu}{\delta A_l^\mu + \delta A_h^\mu}, \qquad (3)$$

between 0 and 1, quantifies by how much the $\delta$-function weight is larger than the low frequency missing area. Basov found $V_c \sim 0.5$ for slightly underdoped $La_{2-x}Sr_xCuO_4$. Figure 3 shows our calculated values in the clean limit, which are much smaller than the measured one.

However in the presence of disorder only the low frequency missing area is expected to be reduced, and eq. (3) becomes for the c direction

$$V_c = \frac{\delta A_h^c}{\delta A_l^c \times p + \delta A_h^c} = V_c^{clean}(\frac{\lambda_c}{\lambda_c^{clean}})^2 \qquad (4)$$

where the disorder parameter p in the presence of sum rule violation is given by

$$p = (\frac{\lambda_{clean}}{\lambda})^2(1-V_c) = (\frac{\lambda_{clean}}{\lambda})^2 - V_c^{clean} \qquad (5)$$

instead of Eq. (1). We obtain p from the comparison between experimental and calculated penetration depth anisotropies in Fig. 2 and the calculated $V_c^{clean}$, and plot the resulting sum rule violation in the c direction and 1/p in Fig. 4. For slightly underdoped $La_{2-x}Sr_xCuO_4$, $T_C=32K$, we obtain $V_c= 60\%$, somewhat larger than Basov's observations. With a smaller nearest neighbor repulsion the sum rule violation decreases, to 28% for V=0 for that doping.

The in-plane sum rule violation, even though it is bigger than the c-axis clean limit one, is still rather small. It will be challenging to detect it experimentally unless it is possible to induce enough disorder to suppress the in-plane superfluid weight (increase the penetration depth) without causing pairbreaking. The situation is somewhat better for the higher $T_c$ materials such as $Tl_2Ba_2CuO_y$[4]. Still, the in-plane kinetic energy lowering that manifests itself in the small optical sum-rule violation seen in

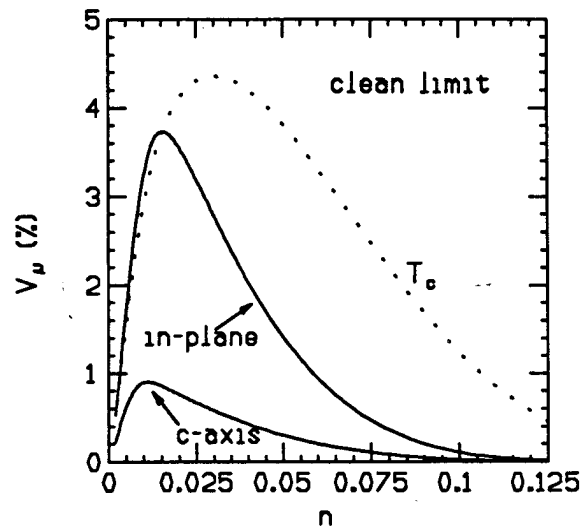

Fig.3: Sum rule violation ($V_\mu$ of Eq. (3)) in the clean limit.

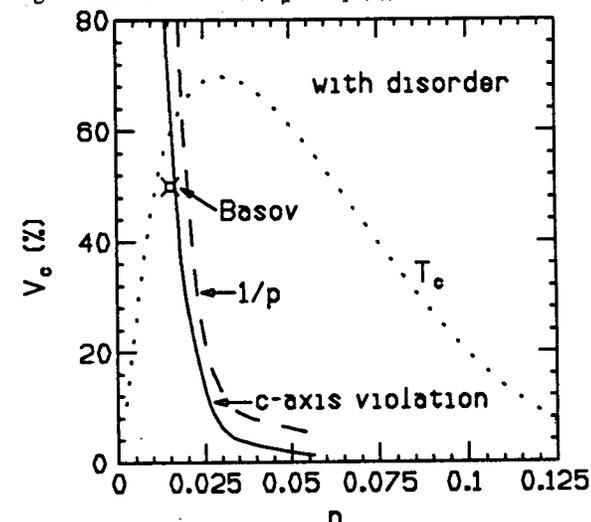

Fig. 4: c-axis sum rule violation (Eq. 4), and disorder parameter $p^{-1}$ (Eq. 5, dashed line) versus hole doping.

Fig. 3 accounts for almost the entire condensation energy of the superconductor within our model[4].